%% file: sm2013-top.tex
\documentclass[12pt]{article}
\usepackage{fancyhdr}
\usepackage{authblk}
\usepackage{feynmf}
\usepackage{amsmath,amssymb}
\usepackage{graphicx}
\usepackage{amssymb}
\usepackage{url,color}
\usepackage{epstopdf}
\usepackage{caption}
\usepackage{subfig}
\usepackage[colorlinks=true, citecolor=green, urlcolor=cyan]{hyperref} 
\def\stacksymbols #1#2#3#4{\def\theguybelow{#2}
    \def\vp{\lower#3pt}
    \def\sp{\baselineskip0pt\lineskip#4pt}
    \mathrel{\mathpalette\intermediary#1}}

\def\intermediary#1#2{\vp\vbox{\sp
     \everycr={}\tabskip0pt
     \halign{$\mathsurround0pt#1\hfil##\hfil$\crcr#2\crcr
              \theguybelow\crcr}}}

\def\beq{\begin{equation}}
\def\eeq#1{\label{#1}\end{equation}}
\def\eeqn{\end{equation}}
\newcommand{\gsim}{\hbox{ \raise3pt\hbox to 0pt{$>$}\raise-3pt\hbox{$\sim$} }}
\newcommand{\lsim}{\hbox{ \raise3pt\hbox to 0pt{$<$}\raise-3pt\hbox{$\sim$} }}

\newenvironment{Eqnarray}%
   {\arraycolsep 0.14em\begin{eqnarray}}{\end{eqnarray}}
\def\beqa{\begin{Eqnarray}}
\def\eeqa#1{\label{#1}\end{Eqnarray}}
\def\eeqan{\end{Eqnarray}}

\def\leqn#1{(\ref{#1})}

\def\invfb{ \mbox{fb}^{-1} }

\def\roots{ \sqrt{s} }
\def\TeV{ \mbox{TeV}}
\def\GeV{ \mbox{GeV}}

\def\be{\begin{equation}}  
\def\ee{\end{equation}}  
\def\bea{\begin{eqnarray}}  
\def\eea{\end{eqnarray}}

\providecommand{\ee}   {\rm{e^+e^-}}

\newcommand{\ttbar}{ t \bar t}

\newcommand{\eplus}{e^+}
\newcommand{\eminus}{e^-}
\newcommand{\epem}{\eplus\eminus}
\def\invfb{ \mbox{fb}^{-1} } 

\newcommand{\afb}{A_{FB}}
\newcommand{\afbt}{A^t_{FB}}

\newcommand{\cthel}{\mathrm{cos} \theta_{hel}}

\newcommand{\Zzero}{Z^0}
\newcommand{\tpq}{t}

\input mymacros.tex

\title{ \LARGE\bf Top quark precision physics at the International Linear Collider\footnote{This is a shortened
version of the physics book for the ILC published in~\cite{Baer:2013cma}. Where applicable updates of the results 
have been integrated.}}
\author[1]{D.~Asner}
\author[2]{A.~Hoang}
\author[3]{Y.~Kiyo}
\author[4]{R.~P\"oschl\thanks{Corresponding author: poeschl@lal.in2p3.fr}}
\author[5]{Y.~Sumino}
\author[6]{M.~Vos}
\affil[1]{\footnotesize Pacific Northwest National Laboratory 902 Battelle Boulevard Richland, WA, USA}
\affil[2]{\footnotesize Inst. f\"ur Theor. Physik, Universit\"at Wien, Boltzmanngasse 5, A-1090 Vienna, AUSTRIA}
\affil[3]{\footnotesize Department of Physics, Juntendo University, Inzai, Chiba 270-1695, JAPAN}
\affil[4]{\footnotesize LAL, CNRS/IN2P3, Universit\'e Paris Sud, F-91898 Orsay CEDEX, FRANCE}
\affil[5] {\footnotesize Graduate School of Science,Tohoku University, 6-3,Aramaki Aza-Aoba,Aoba-ku,Sendai 980-8578, JAPAN}
\affil[6] {\footnotesize IFIC, Universitat de Valencia CSIC, c/ Catedr\`atico Jos\'e Beltr\`an, 2  46980 Paterna, SPAIN}

\begin{document}

\date{}

\maketitle
\thispagestyle{fancy}

\begin{abstract}
Top quark production in the process $e^+e^- \rightarrow t\bar{t}$ at a future linear electron positron collider with polarised beams is a powerful tool to determine the scale of new physics.  
Studies at the $\ttbar$ threshold will allow for precise determination of the top quark mass in a well defined theoretical framework. At higher energies vector, axial vector and tensorial $CP$ conserving couplings can be separately determined for the photon and the $Z^0$ component in the electroweak production process. 
The sensitivity to new physics  would be dramatically improved w.r.t. to what expected from LHC for electroweak couplings. 
\end{abstract}












\def\thefootnote{\fnsymbol{footnote}}
\setcounter{footnote}{0}
%

\input intro.tex
\input top-mass.tex
\input ilc-coupl.tex

\input top-remarks.tex
\bibliographystyle{utphys_mod}
\begin{footnotesize}
\bibliography{sm2013-top}
\end{footnotesize}

\end{document}

%% file: mymacros.tex
\let\bar=\overbar

\def\lsim{\mathrel{\raise.3ex\hbox{$<$\kern-.75em\lower1ex\hbox{$\sim$}}}}
\def\gsim{\mathrel{\raise.3ex\hbox{$>$\kern-.75em\lower1ex\hbox{$\sim$}}}}

\def\del{\partial}
\def\Dslash{\not{\hbox{\kern-4pt $D$}}}
\def\dslash{\not{\hbox{\kern-2pt $\del$}}}

\def\ee{e^+e^-}

\def\alphas{\alpha_s}
\def\msb{{\bar{\scriptsize M \kern -1pt S}}}

\def\drb{{\bar{\scriptsize D \kern -1pt R}}}

\makeatletter
\def\section{\@startsection{section}{0}{\z@}{5.5ex plus .5ex minus
 1.5ex}{2.3ex plus .2ex}{\large\bf}}
\def\subsection{\@startsection{subsection}{1}{\z@}{3.5ex plus .5ex minus
 1.5ex}{1.3ex plus .2ex}{\normalsize\bf}}
\def\subsubsection{\@startsection{subsubsection}{2}{\z@}{-3.5ex plus
-1ex minus  -.2ex}{2.3ex plus .2ex}{\normalsize\sl}}

\renewcommand{\@makecaption}[2]{%
   \vskip 10pt
   \setbox\@tempboxa\hbox{\small #1: #2}
   \ifdim \wd\@tempboxa >\hsize     
       \small #1: #2\par          
     \else                        
       \hbox to\hsize{\hfil\box\@tempboxa\hfil}
   \fi}

 \def\citenum#1{{\def\@cite##1##2{##1}\cite{#1}}}
 
\newcount\@tempcntc
\def\@citex[#1]#2{\if@filesw\immediate\write\@auxout{\string\citation{#2}}\fi
  \@tempcnta\z@\@tempcntb\m@ne\def\@citea{}\@cite{\@for\@citeb:=#2\do
    {\@ifundefined
       {b@\@citeb}{\@citeo\@tempcntb\m@ne\@citea\def\@citea{,}{\bf ?}\@warning
       {Citation `\@citeb' on page \thepage \space undefined}}%
    {\setbox\z@\hbox{\global\@tempcntc0\csname b@\@citeb\endcsname\relax}%
     \ifnum\@tempcntc=\z@ \@citeo\@tempcntb\m@ne
       \@citea\def\@citea{,}\hbox{\csname b@\@citeb\endcsname}%
     \else
      \advance\@tempcntb\@ne
      \ifnum\@tempcntb=\@tempcntc
      \else\advance\@tempcntb\m@ne\@citeo
      \@tempcnta\@tempcntc\@tempcntb\@tempcntc\fi\fi}}\@citeo}{#1}}
\def\@citeo{\ifnum\@tempcnta>\@tempcntb\else\@citea\def\@citea{,}%
  \ifnum\@tempcnta=\@tempcntb\the\@tempcnta\else
  {\advance\@tempcnta\@ne\ifnum\@tempcnta=\@tempcntb \else\def\@citea{--}\fi
    \advance\@tempcnta\m@ne\the\@tempcnta\@citea\the\@tempcntb}\fi\fi}
\makeatother

%% file: intro.tex
\section{Introduction}

The top quark, or $t$~quark, is by far the heaviest particle of the 
Standard Model. Its large mass implies that this is the Standard Model 
particle most strongly coupled to the mechanism of electro-weak symmetry 
breaking.  For this and other reasons, the 
$t$~quark is expected to be a window to any new physics at the TeV 
energy scale.   In this section, we will review the ways that new
 physics might appear in the precision study of the $t$~quark and the 
capabilities of the ILC to discover these effects.

The $t$~quark was discovered at the Tevatron proton-antiproton collider 
by the D0 and CDF 
experiments~\cite{Abe:1995hr,Abachi:1995iq}.  Up to now, the $t$~quark has only 
been studied at hadron colliders, at the Tevatron and, 
only in past two years,  at the LHC. The Tevatron experiments accumulated
 a data sample of about $12\,\invfb$ 
in Run I and Run II, at center of mass energies of 
$1.8\,\TeV$ and $1.96\,\TeV$, respectively. About half of this data
is fully analyzed. At the LHC, a data sample of about $5\,\invfb$ 
has been recorded at a center-of-mass energy of $7\,\TeV$ up to the 
end of 2011.   In 2012, the machine has operated at a center of mass
energy of $8\,\TeV$.  A non exhaustive collection of results obtained at hadron colliders is given in Table~\ref{tab:hadsum}.
 
\begin{table}[t]
\begin{center}
\begin{scriptsize}
\begin{tabular}{|ccc|}
\hline
Observable & Experiment & Value \\
\hline \hline
Cross section $\sigma_{\ttbar}$ & CDF~\cite{Aaltonen:2010ic} &$\mathrm{7.82\pm0.38\,(stat.)\,\pm0.37\,(syst.)\,\pm 0.15\,(th.)\,pb}$ \\
                                & D0~\cite{Abazov:2011mi} &$\mathrm{7.78\pm^{+0.77}_{-0.64}\,(stat+syst.+lumi)\,pb}$ \\
                                & ATLAS 7\,TeV~\cite{ATLAS-CONF-2012-024} &$\mathrm{177\pm3\,(stat.)\,^{+8}_{-7}\,(syst.)\,\pm7\,(lumi.)\,pb}$ \\
                                & ATLAS 8\,TeV~\cite{ATLAS-CONF-2012-149} &$\mathrm{241\pm2\,(stat.)\,\pm31\,(syst.)\,\pm9\,(lumi.)\,pb}$ \\ 
                                & CMS 7\,TeV~\cite{Chatrchyan:2012bra}  &$\mathrm{161.9 \pm2.5\,(stat.)\,^{+5.1}_{-5.0}\,(syst.)\,\pm3.6\,(lumi.)\,pb}$ \\
                                & CMS 8\,TeV~\cite{CMS-PAS-TOP-12-007}  &$\mathrm{227 \pm3\,(stat.)\,\pm11\,(syst.)\,\pm10\,(lumi.)\,pb}$ \\
\hline
Top quark mass $m_t$            &  TEVATRON~\cite{CDF:2013jga}                 & $\mathrm{173.2\pm0.51\,(stat.)\,\pm\,0.71 (syst.)\,\GeV} $ \\ 
                                &  LHC~\cite{ATLAS:2012coa} & $\mathrm{173.3\pm0.5\,(stat.)\,\pm\,1.3 (syst.)\,\GeV} $\\

\hline
Top quark width $\Gamma_t$ & CDF~\cite{bib:cdf-gt2012} & $\mathrm{2.21^{+1.84}_{-1.11}\,\GeV} $ \\
                           & D0~\cite{Abazov:2010tm}& $\mathrm{1.99^{+0.69}_{-0.55}\,\GeV} $ \\
\hline
$BR(t\rightarrow W b)/BR(t \rightarrow W q)$ & CDF~\cite{bib:cdf-r2012}& $0.94\pm0.09$ \\ 
                                             & D0~\cite{Abazov:2011zk}& $0.9\pm0.04$\\ 
                                             & CMS~\cite{CMS-PAS-TOP-12-035}& $\mathrm{1.023^{+0.036}_{-0.034}}$ \\
\hline
Forward backward asymmetry $\afbt$ & CDF~\cite{bib:cdf-afb2012} & $\mathrm{0.162\pm0.047\,(stat.+syst.)} $ \\
                                   & D0~\cite{Abazov:2011rq} & $0.196\pm0.060\,\mathrm{(stat.)} ^{+0.018}_{-0.026}\,\mathrm{(syst.)}$ \\
\hline
Charge asymmetry $A_C$ & ATLAS~\cite{ATLAS:2012an}& $\mathrm{-0.018\pm0.028\,(stat.)\,\pm\,0.023 (syst.)}$ \\
                       & CMS~\cite{Chatrchyan:2011hk}& $-0.013\pm0.028\,\mathrm{(stat.)}^{+0.029}_{-0.031}\,\mathrm{(syst.)}$ \\
\hline
$W$~boson helicity $f_0$ & CDF~\cite{Aaltonen:2012lua}& $\mathrm{0.726\pm0.066\,(stat.)\,\pm\,0.067 (syst.)}$ \\
                         & ATLAS~\cite{Aad:2012ky}& $\mathrm{0.67\pm0.03\,(stat.)\,\pm\,0.06 (syst.)}$ \\
                         & CMS~\cite{CMS-PAS-TOP-12-015}& $\mathrm{0.698\pm0.057\,(stat.)\,\pm\,0.064 (syst.)}$ \\
\hline

\hline

\end{tabular}
\end{scriptsize}
\caption{Collection of recent results on $\ttbar$ production obtained at hadron colliders. The table reflects the status of the Winter conferences 2013.}
\label{tab:hadsum}
\end{center}
\end{table}


The ILC would be the first machine at which the $t$~quark is studied
 using a precisely defined leptonic initial state.  This brings
the $t$~quark into an environment in which individual events can be 
analyzed in more detail, as we have explained in the Introduction. 
It also changes the production mechanism  for $t$~ quark pairs from 
the strong to the electro-weak interactions, which are a step closer 
to the phenomena of electro-weak symmetry breaking that we aim to 
explore.  Finally, this change brings
into play new experimental observables---weak interaction 
polarization and parity asymmetries---that are very sensitive 
to the coupling of the $t$~quark to possible new interactions.   
It is very possible that, while the $t$~quark might respect Standard 
Model expectations at the LHC, it will break those expectations 
when studied at the ILC.

%% file: top-mass.tex
\section{$\ee \rightarrow \ttbar$ at threshold}
One of the unique capabilities of an $\ee$ linear collider is the ability
to carry out cross section measurements at particle 
production thresholds.  The accurately known and readily variable beam 
energy of the ILC makes it possible to measure the shape of the cross section
at any pair-production threshold within its range.
Because of the leptonic initial state, it is also  possible to tune 
the initial spin state, giving additional options for precision threshold 
measurements. 
The $t\bar t$ pair production threshold, located at a center of mass energy
energy  $\sqrt{s}\approx 2 m_t$,  allows for precise measurements of 
the $t$~quark mass $m_t$ as well as the $t$~quark 
total width $\Gamma_t$ and the QCD coupling $\alpha_s$. 
 Because the top is a spin-$\frac{1}{2}$ fermion, the
 $t\bar t$ pair is produced in an angular $S$-wave state.  This leads to
a clearly visible rise of the cross section even when folded with the
 ILC luminosity spectrum. Moreover, because the top pair is 
produced in a color singlet state, the experimental measurements can be 
compared with very accurate and unambiguous 
analytic theoretical predictions of the cross section 
with negligible hadronization effects. The dependence of the 
$t$~quark cross section shape on the $t$~quark mass and interactions
is computable to high precision with full control over the renormalization 
scheme dependence of the top mass parameter.  In this section, we will 
review the expectations for the theory and ILC measurements of the 
$t$~quark threshold cross section shape. The case of the $t$~quark
threshold is not only important in its own right but also serves
 as a prototype case for other particle thresholds that might be accessible at the ILC.

\subsection{ Status of QCD theory}

\begin{figure}
\centering
\includegraphics[width=0.7\columnwidth]{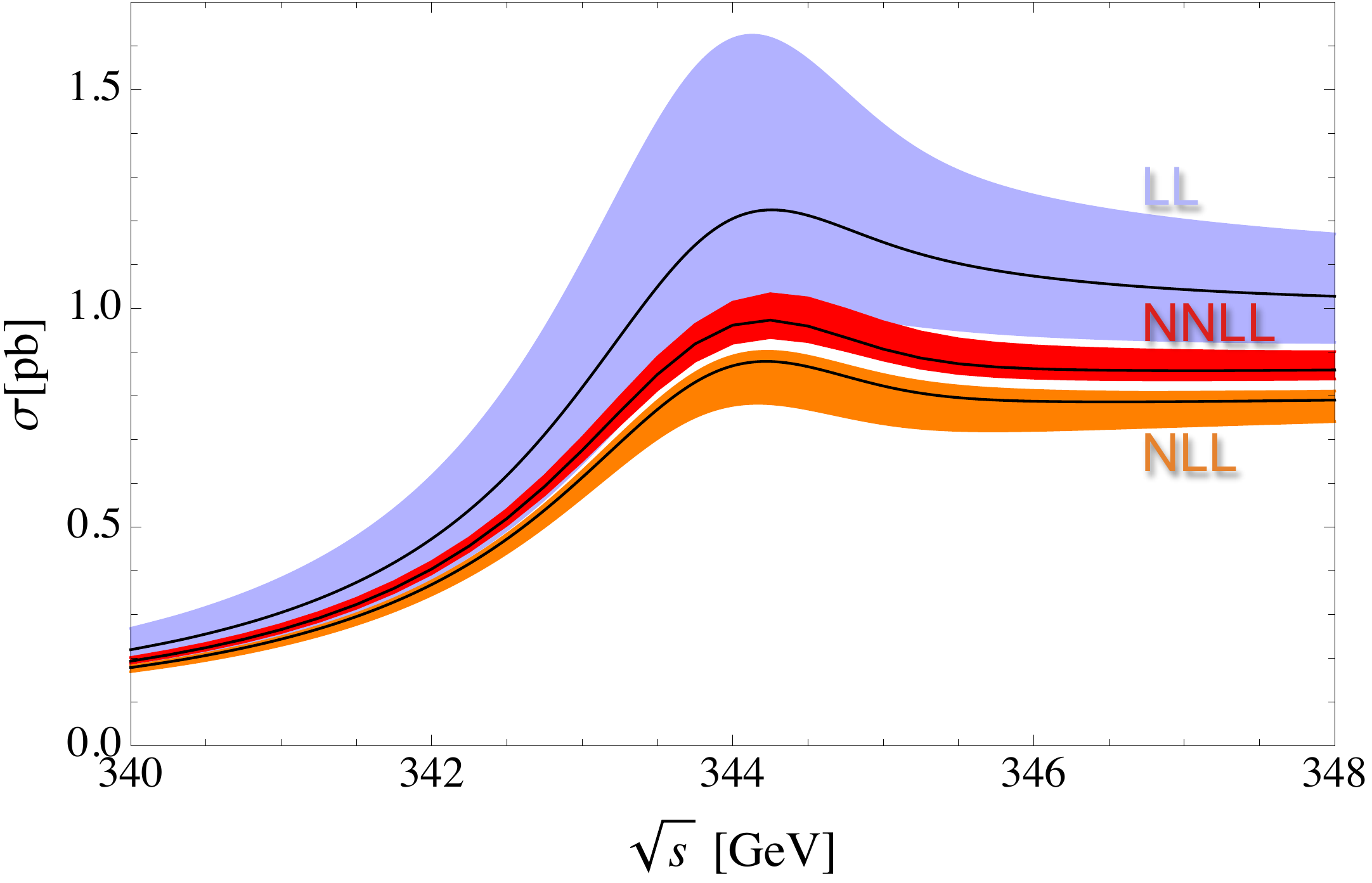}
\caption{
Bands show the theoretical QCD uncertainties of the prediction of the top pair production cross section at the $t\bar t$ threshold
   at the ILC as achieved by recent renormalization-group-improved QCD calculations~\cite{Hoang:2013uda}. Their analysis estimated 
   the theoretical uncertainty in the total cross section as $d\sigma_{t\bar t}/\sigma_{t\bar t}\pm 5\%$. For further explanation see text.
}
\label{fig:tthresh-hs}
\end{figure}

The calculation of the total top pair production cross section makes use of the 
 method of non-relativistic effective 
theories. The $t$~quark mass parameter used in this calculation is
defined at the scale of about 10~GeV corresponding to the typical physical
separation of the $t$ and $\bar t$.  This mass parameter can be
converted to the $\msb$ mass in a controlled way. The summation of QCD
 Coulomb singularities treated by a 
non-relativistic fixed-order expansion is well known up to 
NNLO~\cite{Hoang:2000yr} and has recently been extended accounting 
also for NNNLO corrections~\cite{Beneke:2008cr}. Large QCD velocity 
logarithms have been determined using renormalization-group-improved 
non-relativistic perturbation theory up to NLL order, with a partial 
treatment of NNLL effects~\cite{Hoang:2001mm,Hoang:2003ns,Pineda:2006ri}.
Recently the dominant ultrasoft NNLL corrections have been 
completed~\cite{Hoang:2006ht,Pineda:2011aw,Hoang:2011gy}. The accuracy
in this calculation is illustrated in Fig.~\ref{fig:tthresh-hs}.

Since the $t$~quark kinetic energy is of the order of the $t$~quark width, 
electro-weak effects, which also include finite-lifetime and interference 
contributions, are crucial as well. This makes the cross section dependent 
on the experimental prescription concerning the reconstructed final state. 
Recently the NLO non-resonant calculation of $\ttbar$ production in~\cite{Beneke:2010mp,Penin:2011gg} has been extended to NNLO accuracy~\cite{Hoang:2010gu,Jantzen:2013gpa}.

Theoretical predictions for differential cross sections such as the 
top momentum distribution and forward-backward 
asymmetries are only known at the NNLO level 
and are thus much less developed.

\begin{figure}
\centering
\includegraphics[width=0.50\columnwidth]{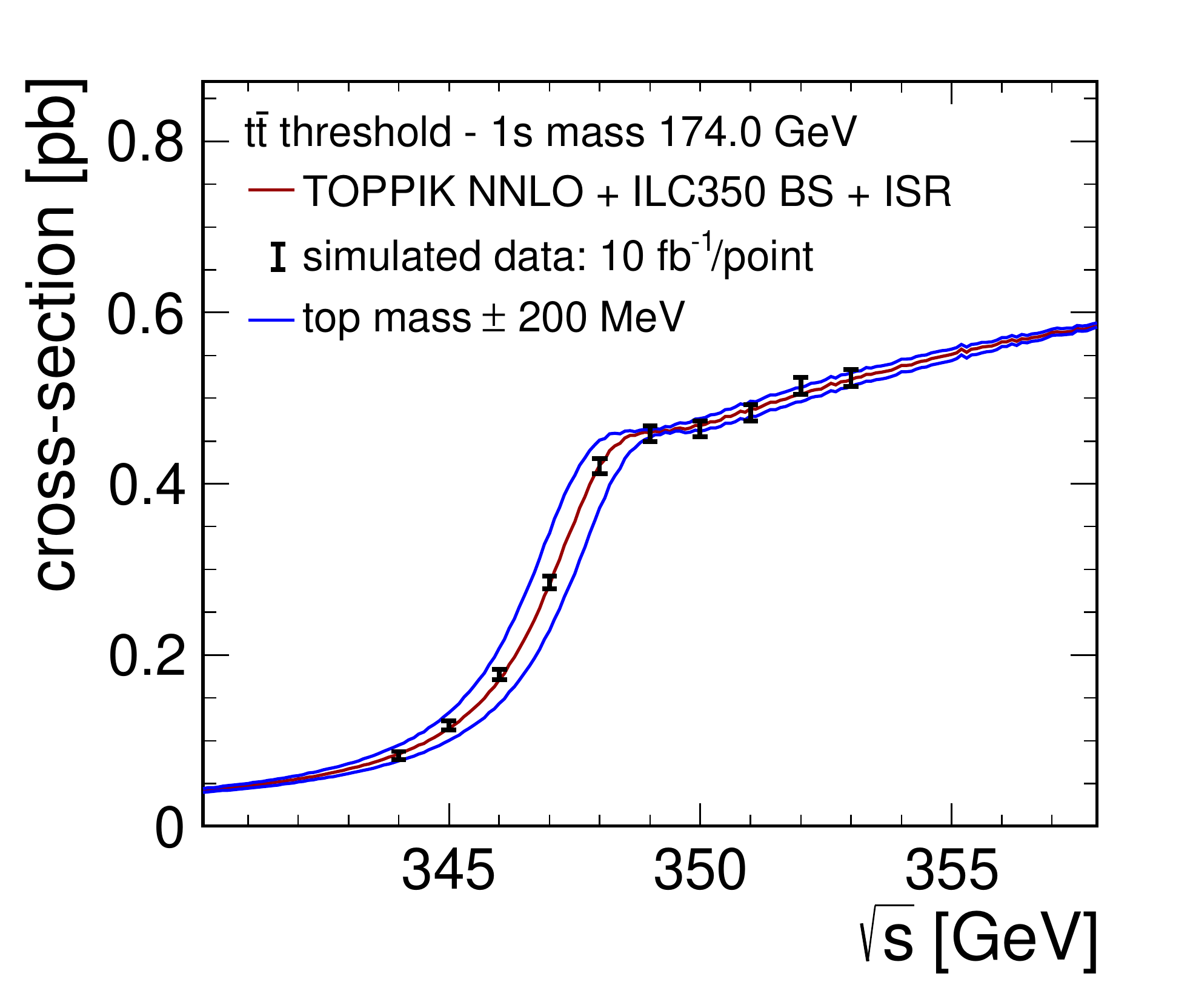}
\includegraphics[width=0.40\columnwidth]{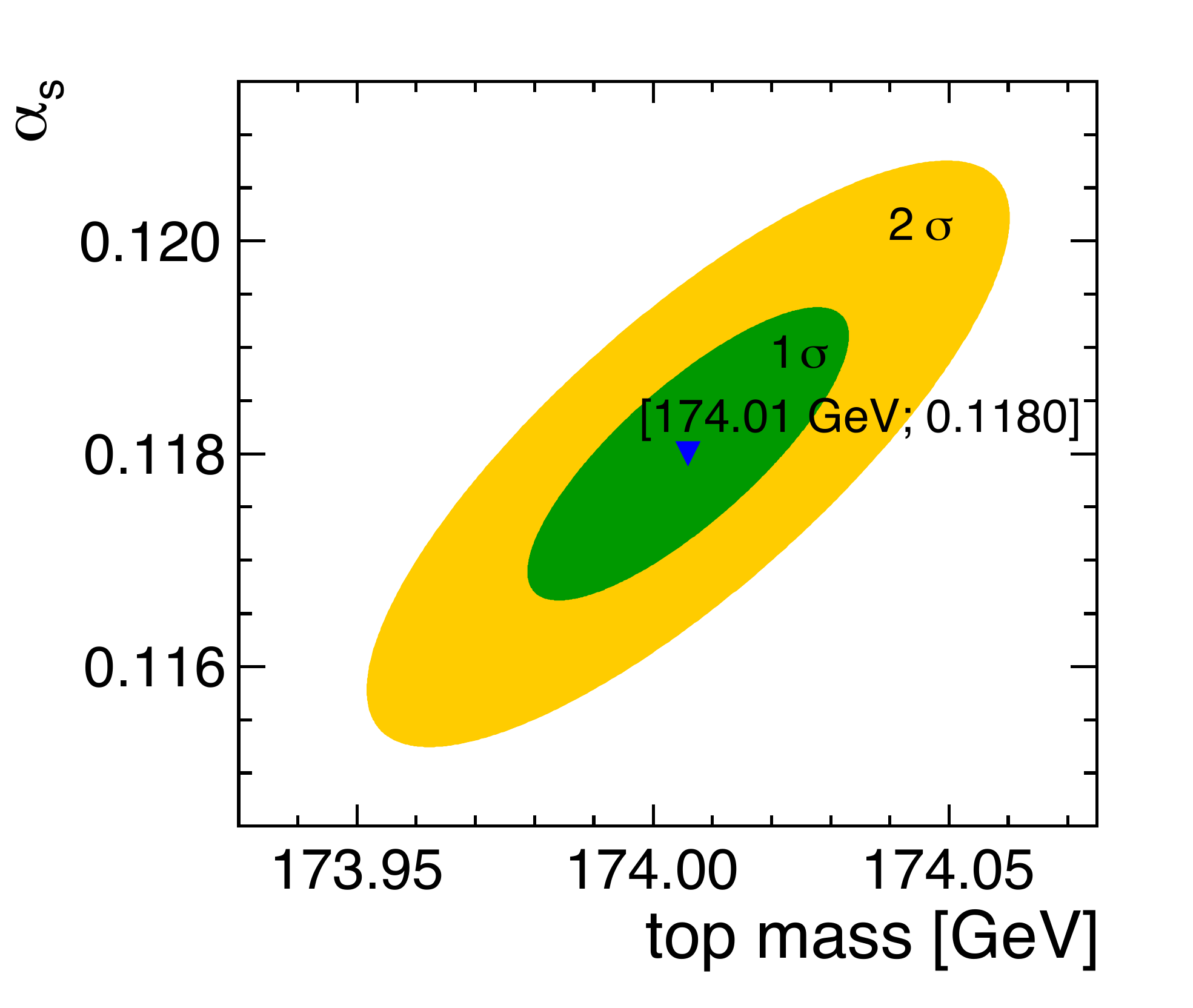}
\caption{Illustration of a $t$~quark threshold measurement at the ILC. 
In the simulation, the $t$~quark mass has been 
chosen to be 174.~GeV.  The blue lines show the effect of varying this 
mass by 200~MeV. The study is based on full detector simulation and takes 
initial state radiation (ISR) and beamstrahlung (BS) and other relevant 
machine effects into account: (left) the simulated threshold scan.  (right) 
error ellipse for the determination of $m_t$ and $\alpha_s$. The figure is
 taken from~\cite{Seidel:2013sqa}.}
\label{fig:ttthresh-scan}
\end{figure}
 
\subsubsection{Simulations and measurements}

The most recent experimental study has been carried out by Seidel, Simon, and Tesar, for which the 
results are shown in Fig.~\ref{fig:ttthresh-scan}~\cite{Seidel:2013sqa}. That study is based on a full detector simulation 
using the ILD detector. It takes the initial state radiation and beamstrahlung of the colliding beams into 
account. has been carried out by Martinez and Miquel in~\cite{Martinez:2002st}.
These authors assumed a total integrated luminosity of $100$~fb$^{-1}$, distributed over 10 equidistant energy points 
in a $10$~GeV range around the threshold, using the ILC and CLIC beam parameters. To treat the 
strong correlation of the input theory parameters, simultaneous fits 
were carried out for the $t$~quark mass, the QCD coupling and the $t$~quark width from 
measurements of the total cross section that was simulated based on the code TOPPIK with NNLO corrections~\cite{Hoang:1999zc}. 
The results are shown in Table~\ref{tab:ThresholdResults} and demonstrate that the statistical precision of the $t$~quark mass in this study is of the order of 30\,MeV.
The results are compatible with the results reported in~\cite{Horiguchi:2013wra}. There the width of the $t$ quark is determined to a statistical precision at the 2\% level.   

\begin{table}
\centering
\begin{tabular}{|ccc|}
\hline
\multicolumn{3}{|c|}{1S top mass and $\alpha_s$ combined 2D fit}\\
\hline
Error type & ILC & CLIC \\
\hline
 $m_t$ stat. error &  27~MeV & 34~MeV\\
 $m_t$ theory syst. (1\%/3\%) &  5~MeV / 9~MeV & 5~MeV / 8~MeV\\
 $\alpha_s$ stat. error & 0.0008 & 0.0009\\
 $\alpha_s$ theory syst. (1\%/3\%) & 0.0007 / 0.0022 & 0.0008 / 0.0022 \\
\hline
\end{tabular}

\caption{Results summary for the 2D simultaneous top mass and $\alpha_s$ determination with a threshold scan at CLIC and ILC for 10 points with a total integrated luminosity of 100\,fb$^{-1}$~\cite{Seidel:2013sqa}. \label{tab:ThresholdResults}}
\end{table}





As shown in an older study by Martinez and Miquel in~\cite{Martinez:2002st} even better precision on the $t$ mass may be achieved by taking 
other observables into account.

The threshold $t$~quark mass determined in the study of Seidel et al. must still be 
converted to the standard $t$~quark $\msb$ mass.  The conversion 
formula, to three-loop order, is given in \cite{Hoang:1999zc}.  The
conversion adds an error of about 100~MeV from truncation of the QCD
perturbation series and an error of 70~MeV for each uncertainty of 
$0.001$ in the value of $\alphas$. Both sources of uncertainty should be 
reduced by the time of the ILC running.  In particular, the study of 
event shapes in $\ee\to q\bar q$ at the high energies available at
ILC should resolve current questions concerning the precision determination of
$\alphas$.  We recall that these estimates are the results of a 
precision theory of the relation between the threshold mass and the 
$t$~quark $\msb$ mass. A comparable theory does not yet exist for the
conversion of the $t$~quark mass measured in hadronic collisions to the
$\msb$ value.

The precise determination of the $t$~quark mass is likely to have
important implications for fundamental theory. A value of the top
quark mass accurate at the level that a linear collider will provide is for example a key input to models of the 
vacuum stability of the universe.


%% file: ilc-coupl.tex
\section{Probing the top quark vertices at the ILC}~\label{sec:top-elweak}

At higher energy, the study of $t\bar t$ pair production concentrates on the precise measurements of the couplings
of the $t$~quark to the $Z^0$~boson and the photon. 
 In contrast to the situation  at hadron colliders, the leading-order pair production process
$\ee\to t\bar t$ goes directly through the $t\bar{t} Z^0$ and 
$t\bar{t} \gamma$ vertices.  There is no concurrent QCD production 
of $\ttbar$ pairs, which increases greatly the potential for a clean measurement.
A commonly used expression to describe the  the current at the $t\bar{t} X$ vertex is~\cite{Juste:2006sv}

\begin{equation}\label{eq:gordon}
\Gamma_\mu^{ttX}(k^2,\,q,\,\bar{q}) = ie \left\{
  \gamma_\mu \, \left(  \widetilde{F}_{1V}^X(k^2)
                      + \gamma_5\widetilde{F}_{1A}^X(k^2) \right)
+ \frac{(q-\bar{q})_\mu}{2m_t}
    \left(  \widetilde{F}_{2V}^X(k^2)
          + \gamma_5\widetilde F_{2A}^X(k^2) \right)
\right\} .
\end{equation}
where $X = \gamma,Z$ and the  $\widetilde{F}$ are related to the usual
form factors $F_1$, $F_2$ by 
\begin{equation}
\label{eq:rel1}
\widetilde F^X_{1V} = -\left( F^X_{1V}+F^X_{2V} \right) \, , \qquad
\widetilde F^X_{2V}  =  F^X_{2V} \, , \qquad
\widetilde F^X_{1A} = -F^X_{1A} \, , \qquad
\widetilde F^X_{2A} =  -iF^X_{2A} \, .
\end{equation}
In the Standard Model the only form factors which are different 
from zero are $F_{1V}^\gamma(k^2),\,F_{1V}^{Z}(k^2)$ and $F_{1A}^{Z}(k^2)$.  The
quantities $F_{2V}^{\gamma,Z}(k^2)$ are the electric and weak magnetic dipole 
moment (EDM and MDM) form factors. The presence of the $\gamma/Z^0$ interference in electro-weak production gives sensitivity to the actual sign of the coupling constants. This is a distinct difference to the associated vector boson production at the LHC, which is only sensitive to their absolute values.

In the following section, we will review the importance of measuring these couplings precisely.   Then we will describe studies of the experimental  capabilities of the ILC to perform these measurements. A great asset to test the chiral structure is the availability of polarized beams at the ILC. The studies presented in the following exploit this potential. For a full overview on the physics potential with polarised beams the reader is referred to~\cite{MoortgatPick:2005cw}.

\subsection{Top quark and new physics -  A brief motivation}

Neither the Standard Model of particle physics nor its super-symmetric extensions deliver an explanation for the striking mass hierarchy in the fermion sector. On the other hand, the mass hierarchy can be accommodated in models featuring extra dimensions, which are dual to models in which the $t$~quark and the Higgs Boson are composited objects. 
An example model with extra dimensions is that by Randall and Sundrum~\cite{Randall:1999ee}.  New physics will modify the electro-weak $\Zzero \ttbar$ vertex and modify the couplings $Q_L$ and $Q_R$ to the left- and right-chiral parts of the $\tpq$-quark wave-function. New physics may also entail the existence of a new $Z^{0'}$~boson or Kaluza-Klein excitations of the Standard Model $\Zzero$~boson. The modified vertex gives rise to different forward backward asymmetries $\afb$ than those predicted by the Standard Model. For this quantity a 3$\sigma$ discrepancy of the effective electro-weak mixing angle ${\rm sin}^{2}\theta_{eff}$ obtained in $b\bar{b}$ production at LEP has yet to be resolved~\cite{ALEPH:2005ab}. If this  effect is real, it is likely to be larger for the heavy $t$~quark.  
Please note that tensions with the Standard Model are also reported for the Tevatron results on $\afb$ in $\ttbar$ final states. 
The Standard Model prediction of the weak mixing angle is also challenged by the about  3$\sigma$ discrepancy observed in the
left right asymmetry $A_{LR}$ at the linear collider SLC, which featured linearly polarized electron beams. This measurement is the most precise measurement of $sin^{2}\theta_{eff}$ as of today. 

\subsection{ILC measurements}

In the previous section, theories with extra dimensions and/or in which the 
$t$~quark and the Higgs boson are composite were briefly introduced. Compositeness is an
essential element of the physics of electro-weak symmetry breaking.
A key test of this idea would come from the measurement of the $t\bar
t Z$ couplings, where significant deviations from the predictions of
the Standard Model would be expected. The ILC provides an ideal environment to measure these couplings. 
At the ILC $\ttbar$ pairs would be copiously produced, several 100,000 events 
for an integrated luminosity of $500\,\invfb$.  The production is by
$s$-channel $\gamma$ and $\Zzero$~boson exchange, so the couplings to the $\Zzero$ enter the
cross section in order 1.  It is possible  to almost entirely eliminate the background from other Standard Model processes.  

With the use of polarized beams, $t$ and $\bar t$ quarks oriented
toward different angular regions in the detector are enriched 
in left-handed or right-handed $t$~quark
polarization~\cite{Parke:1996pr}. 
This means that the experiments can independently access the 
couplings of left- and right-handed polarized quarks to the $\Zzero$~boson.
In principle, the measurement of the  cross section and forward-backward
asymmetry for two different polarization settings measures both the
photon and $\Zzero$~boson couplings of the $t$~quark for each handedness.
New probes of the $t$~quark  decay vertices are also available, although we expect that
these will already be highly constrained by the LHC measurements of
the $W$ polarization in $t$~quark decay.

\begin{figure}
\centering
\includegraphics[width=0.7\columnwidth]{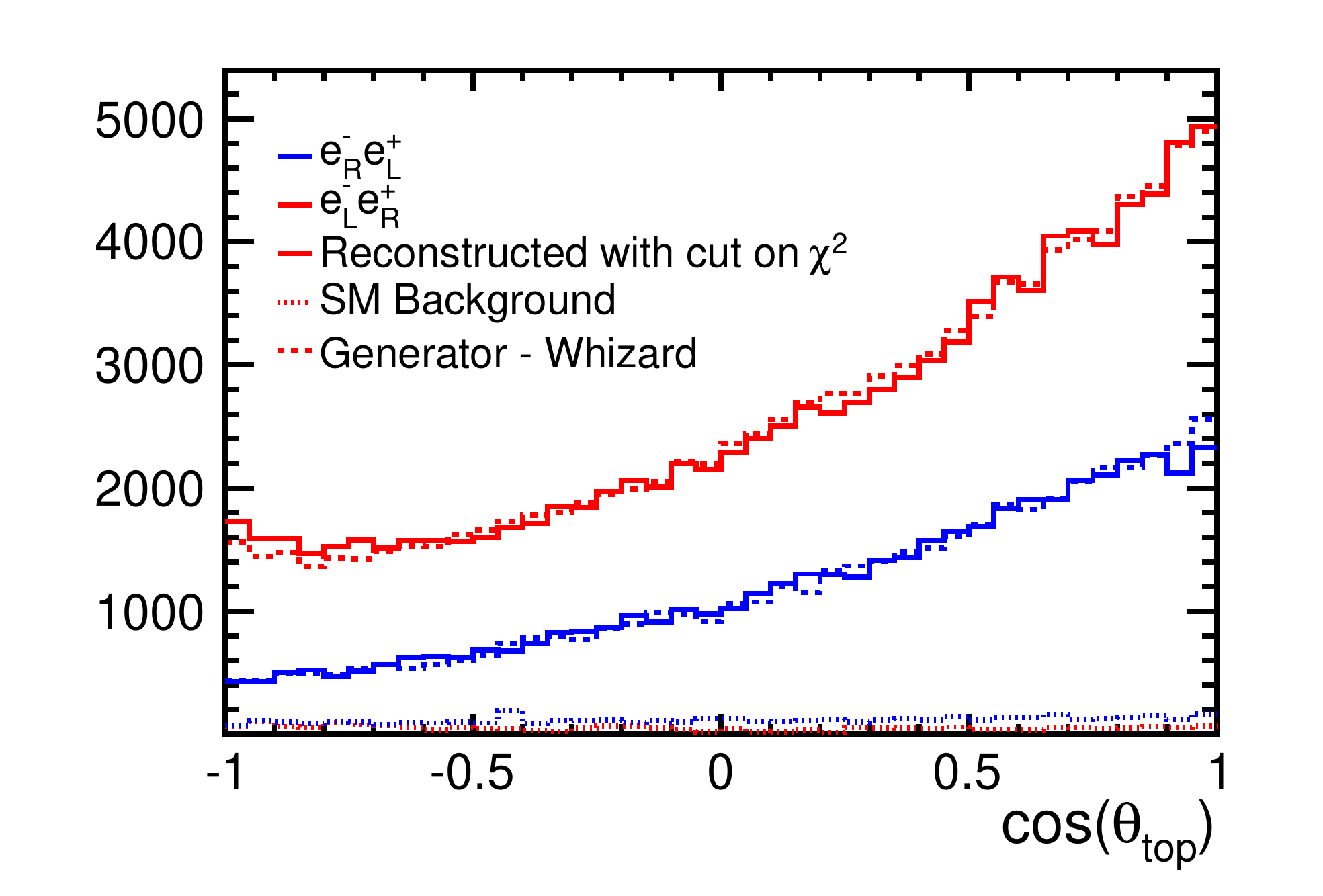}
\caption{Reconstruction of the direction of the $t$ quark for two 
different beam polarization~\cite{Amjad:2013hca}. The population in the two different 
hemispheres w.r.t. the polar angle $\theta_{top}$ allows for the measurement of the forward-backward asymmetry $A_{FB}$. 
The residual Standard Model background is very small.}
\label{fig:topangle}
\end{figure}

Recent studies based on full simulation of ILC detectors for a center of mass
 energy of $\roots=500\,\GeV$ demonstrate that a precision on the 
determination of the couplings the left and the right chiral parts of 
the $t$~quark wave function to the $\Zzero$~boson of up to 1\% can be 
achieved~\cite{ild:bib:benchmark:doublet,Devetak:2010na,Doublet:2012wf}. The most recent example for such a study of semi-leptonic $\ttbar$ decays with full detector simulation is shown in Figure~\ref{fig:topangle}~\cite{Amjad:2013hca}.
The figure demonstrates the clean reconstruction of the $t$~quark direction, which allows for the precise determination of the 
forward-backward asymmetry, which is nearly free of Standard Model background.   It has to be noted however, that the final state gives rise to ambiguities in the correct association 
of the $b$~quarks to the $W$~bosons, see~\cite{Amjad:2013hca} for an explanation. These ambiguities can be nearly eliminated by requiring 
a high quality of the event reconstruction. The control of the ambiguities however requires an excellent detector performance and event reconstruction.
Another solution is the use of the vertex charge to separate the $t$ and $\bar t$~quark decays. It is shown in~\cite{Devetak:2010na} and confirmed in~\cite{bib:ilc-tdr-dbd} that the high efficiency of vertex tagging in the ILC detectors will make this strategy available. 

\begin{figure}
\centering
\includegraphics[width=0.63\columnwidth]{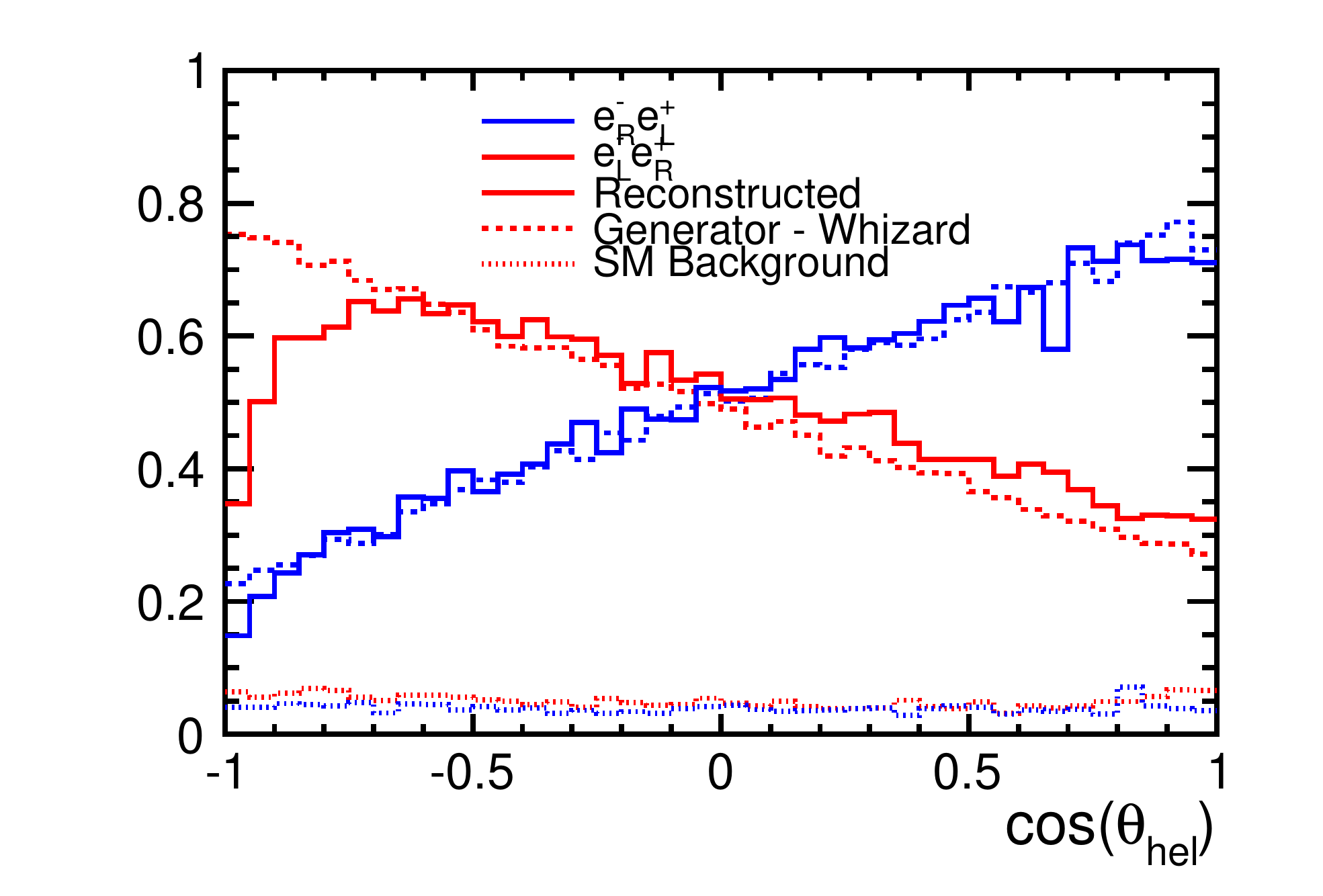}
\includegraphics[width=0.63\columnwidth]{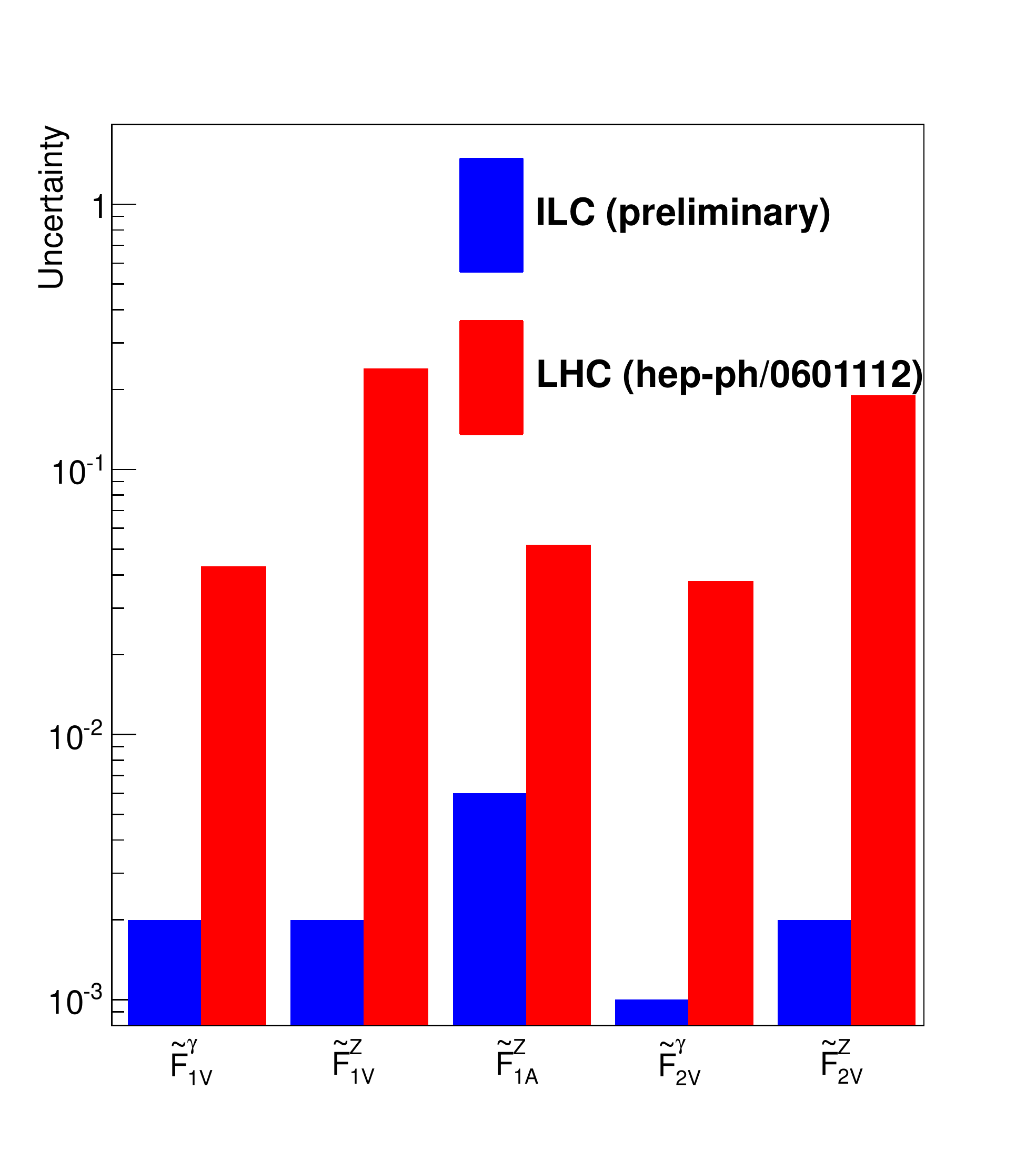}
\caption{
Top: Generated and reconstructed distribution of the helicity angle $\cthel$; 
Bottom: Comparison of precisions on $CP$ conserving form factors  
of the $t$~quark to $\gamma$ and $Z$, $\widetilde F^{\gamma,Z}_{1V,A}$,
expected at the LHC, taken from~\cite{Juste:2006sv}, and at the ILC. 
The LHC results assume an integrated luminosity of $\mathcal{L}=300~\invfb$.
The results for ILC~\cite{Amjad:2013hca} assume an integrated luminosity of $\mathcal{L}=500~\invfb$ at 
$\roots=500\,\GeV$ and a beam polarization  $P_{e^-} =\pm0.8,P_{e^+} = \mp0.3$.}
\label{fig:hel-coupl}
\end{figure}

Even more incisive measurements than presented using optimized 
observables are investigated in~\cite{Amjad:2013hca}. 
These observables are the $\ttbar$ pair production cross-section for left- 
and right-handed polarized beams and the fraction of right-handed ($t_R$) 
and left handed $t$~quarks ($t_L$).
Following a suggestion by~\cite{Berger:2012nw} for the Tevatron, the fraction 
of $t_L$ and $t_R$ in a given sample can be determined with the helicity 
asymmetry.  In the $t$~quark rest frame the distribution of the polar angle 
$\theta_{hel}$ of a decay lepton is 
\beq
\frac{1}{\Gamma} \frac{d\Gamma}{d\cthel} = \frac{1+ a_t \cthel}{2}  
\eeq{costhel}
where $a_t$ varies between $+1$ and $-1$ depending on the 
fraction of right-handed ($t_R$) and left handed $t$~quarks ($t_L$).
 The observable $\cthel$ can easily be measured at the ILC.  
This observable is much less sensitive to ambiguities in the 
event reconstruction than the forward backward asymmetry. The 
slope of the resulting linear distribution provides a very 
robust measure of the net polarization of a $t$~quark sample. The result of a full 
simulation study is shown in the top part of Fig.~\ref{fig:hel-coupl}.
 It is demonstrated that over a range in $\cthel$ the generated 
distribution is retained after event reconstruction. The 
reconstruction is nearly perfect for initial right handed electron beams. Remaining discrepancies in case of
 left handed electron beams can be explained by reconstruction 
inefficiencies for low energetic final state leptons. 

The introduced observables, i.e. $\afb$, cross sections and helicity asymmetry are used to disentangle the coupling of the $t$~quark to the photon and to the $\Zzero$.  
In the bottom part of  Fig.~\ref{fig:hel-coupl}, the precision on the form factors expected 
from the LHC and that from the ILC are compared with each other.
  
Numerical values for the expected accuracies at linear $e^+e^-$ colliders, ILC and earlier on TESLA~\cite{AguilarSaavedra:2001rg}, on seven $t$~quark form factors (due to QED gauge invariance the coupling $\widetilde F^\gamma_{1A}$ is fixed to 0), taken from the studies \cite{Abe:2001nq,AguilarSaavedra:2001rg,Amjad:2013hca},
are given in Tabs.~\ref{tab:tab1} and~\ref{tab:tab2}, along with comparisons to the expectations from 
the LHC experiments. Note in passing, that the replacement of TESLA results of Tab.~\ref{tab:tab2} by an ILC study is in preparation.  

From the comparison of the numbers it is justified to assume that the measurements at an electron positron collider lead to a spectacular improvement and thanks to the $\gamma/Z^0$ interference a $\epem$ collider can fix the sign the form factors. At the LHC the $t$~quark couples either to the photon or to the $Z^0$. In that case the cross section is proportional to e.g. $ (F^Z_{1V})^2 +  (F^Z_{1A})^2$. The precision expected at the LHC cannot exclude a sign flip of neither  $F^Z_{1V}$ nor of $F^Z_{1A}$. On the hand the LEP bounds can exclude a sign flip for $F^Z_{1A}$~\cite{Baur:2004uw}, which renders a much better precision for  $\widetilde F^Z_{1A}$ compared with $\widetilde F^Z_{1V}$. Clearly, the precisions that can be obtained at the LHC are to be revisited in the light of the real LHC data. A first result on associated production of vector boson and $\ttbar$ pairs is published in~\cite{Chatrchyan:2013qca}.

\begin{table}[t]
\begin{center}
\begin{footnotesize}
\begin{tabular}{|cccc|}
\hline
 Coupling & LHC~\protect\cite{Juste:2006sv}  & $e^+e^-$~\protect\cite{Abe:2001nq}&
           $e^+e^-$~\protect\cite{Amjad:2013hca}\\
 & $\mathcal{L}=300~\invfb$ & $P_{e^-}=\pm0.8$ & $\mathcal{L}=500~\invfb$,\, $P_{e^{-,+}} =\pm0.8,\mp0.3$
\\
\hline
$\Delta\widetilde F^\gamma_{1V}$ &
$\begin{matrix} +0.043 \\[-4pt] -0.041\end{matrix}$ &
$\begin{matrix} +0.047 \\[-4pt] -0.047 \end{matrix}$ ,
$\mathcal{L}=200~\invfb$&
$\begin{matrix} +0.002 \\[-4pt] -0.002 \end{matrix}$
\\
$\Delta\widetilde F^Z_{1V}$ &
$\begin{matrix} +0.24 \\[-4pt]  -0.62\end{matrix}$ &
$\begin{matrix} +0.012 \\[-4pt] -0.012\end{matrix}$ , $\mathcal{L}=200~\invfb$ &
$\begin{matrix} +0.002 \\[-4pt] -0.002\end{matrix}$
\\
$\Delta\widetilde F^Z_{1A}$ &
$\begin{matrix} +0.052 \\[-4pt]  -0.060\end{matrix}$ &
$\begin{matrix} +0.013 \\[-4pt] -0.013\end{matrix}$ , $\mathcal{L}=100~\invfb$ &
$\begin{matrix} +0.006 \\[-4pt] -0.006\end{matrix}$
\\
$\Delta\widetilde F^\gamma_{2V}$ &
$\begin{matrix} +0.038 \\[-4pt] -0.035\end{matrix}$ &
$\begin{matrix} +0.038 \\[-4pt] -0.038\end{matrix}$ , $\mathcal{L}=200~\invfb$&
$\begin{matrix} +0.001 \\[-4pt] -0.001\end{matrix}$
\\
$\Delta\widetilde F^Z_{2V}$ &
$\begin{matrix} +0.27 \\[-4pt] -0.19\end{matrix}$ &
$\begin{matrix} +0.009 \\[-4pt] -0.009\end{matrix}$ , $\mathcal{L}=200~\invfb$  &
 $\begin{matrix} +0.002 \\[-4pt] -0.002\end{matrix}$
\\
\hline

\end{tabular}
\end{footnotesize}
\caption{Sensitivities achievable at $68.3\%$ CL for the
CP-conserving $t$~quark form factors $\widetilde F^X_{1V,A}$ and $\widetilde F^X_{2V}$ defined in \leqn{eq:gordon}, at LHC and at
the ILC.  The assumed luminosity samples and, for ILC, beam polarization,
 are indicated.
In the LHC studies and in the study
  \cite{Abe:2001nq}, only one form factor at a time is
allowed to deviate from its SM value.
In study~\cite{Amjad:2013hca} the form factors are allowed to vary independently. }
\label{tab:tab1}
\end{center}
\end{table}

\begin{table}[ht]
\begin{center}
\begin{footnotesize}
\begin{tabular}{|ccc|}
\hline 
 Coupling & LHC~\protect\cite{Juste:2006sv}  &
           $e^+e^-$~\protect\cite{AguilarSaavedra:2001rg}\\
 & $\mathcal{L}=300~\invfb$ & 
$\mathcal{L}=300~\invfb$,\, $P_{e^{-,+}} =-0.8$
\\
\hline
$\Delta {\mbox Re}\, \widetilde F_{2A}^{\gamma}$& $\begin{matrix} +0.17 \\[-4pt] -0.17\end{matrix}$ &$\begin{matrix} +0.007 \\[-4pt] -0.007\end{matrix}$   \\
$\Delta {\mbox Re}\, \widetilde F_{2A}^{Z}$& $\begin{matrix} +0.35 \\[-4pt] -0.35\end{matrix}$ &$\begin{matrix} +0.008 \\[-4pt] -0.008\end{matrix}$  \\
$\Delta {\mbox Im}\, \widetilde F_{2A}^{\gamma}$& $\begin{matrix} +0.17 \\[-4pt] -0.17\end{matrix}$  &$\begin{matrix} +0.008 \\[-4pt] -0.008\end{matrix}$  \\
$\Delta {\mbox Im}\, \widetilde F_{2A}^{Z}$& $\begin{matrix} +0.035 \\[-4pt] -0.035\end{matrix}$ & $\begin{matrix} +0.015 \\[-4pt] -0.015\end{matrix}$ \\
\hline
\end{tabular}
\end{footnotesize}
\caption{Sensitivities achievable at $68.3\%$ CL for the $t$~quark
CP-violating magnetic and electric dipole
form factors 
$\widetilde F^X_{2A}$ defined in \leqn{eq:gordon}, at the  LHC and at 
linear $e^+e^-$ colliders as published in the TESLA TDR.  The assumed luminosity samples and, for TESLA, the beam polarization,
 are indicated. In the LHC studies and in the TESLA studies, only one form factor at a time is 
allowed to deviate from its SM value. } 
\label{tab:tab2}
\end{center}
\end{table}

The expected high precision at a linear $\epem$ collider allow for a profound discussion of effects of new physics. The findings can be confronted with predictions in the framework of Randall-Sundrum models and/or compositeness models such as~\cite{Pomarol:2008bh,Djouadi:2006rk,Hosotani:2005nz,Cui:2010ds,Carena:2006bn,Grojean:2013qca} or Little Higgs models as e.g.~\cite{Berger:2005ht}. All these models entail deviations from the Standard Model values of the $t$~quark couplings to the $Z^0$ boson that will be measurable at the ILC. The interpretation of the results presented in Tables~\ref{tab:tab1} and~\ref{tab:tab2} in terms of the cited and maybe other models is in preparation and left for a future publication. {\em Comments and contributions from theory groups are highly welcome.}


\subsubsection{Theory uncertainties on form factors - A brief outline}

The extraction of form factors requires precise predictions of the inclusive $\ttbar$ production rate and several differential distributions. 
As of today the QCD corrections are known to $N^3LO$ accuracy for the cross section and to $NNLO$ accuracy for $\afbt$. It is shown in~\cite{Amjad:2013hca} and references therein that the uncertainty on these corrections for the cross section are below 1\% with an even smaller scale uncertainty. Uncertainties for $\afbt$ yield about the same values.  

Electro-weak corrections are known to NLO accuracy. The correction for the cross section is about 5\% at $\roots=500\,\GeV$. The correction for $\afbt$ is between 10\% and 15\%~\cite{Fleischer:2003kk,Khiem:2012bp}. It is a major task for the future to estimate the size of the two-loop-correction and ultimately to calculate this contributions.

At this point it should also be noted that all of the studies on form factors are based on the WHIZARD event generator~\cite{Kilian:2007gr,Moretti:2001zz}.  This generator provides the final state in terms of six fermion events.
In particular the $\ttbar$ final state is indistinguishable from the final state of single $t$~quark production which give rise to interference terms. The contribution from single top events can be reduced by adequate kinematic cuts. Still, future studies will have to consider the residual contribution.

%% file: top-remarks.tex
\section{Concluding remarks}
The $t$~quark could be a window to new physics associated with light 
composite Higgs bosons and strong coupling in the Higgs sector.
The key parameters here are the electro-weak couplings of the $t$~quark.  
We have demonstrated that the ILC offers unique capabilities
to access these couplings and measure them to the required high level of
 precision. The mass of the $t$~quark, which is a most important quantitiy
in many theories can be measured model independent to a precision of 
 better than 100~MeV. It has however to be pointed out that 
all of these precision measurements require a superb detector performance and 
event reconstruction. The key requirements are the tagging of final state
$b$ quarks with and efficiency and purity of better than 90\% and jet
 energy reconstruction using particle flow of about 4\% in the entire
 accessible energy range. These requirements are met for the ILC detectors
described in the detector volume of~\cite{bib:ilc-tdr-dbd}. 

On the other hand the physics program require state-of-the art theoretical calculations for the observables. While QCD corrections seem to be largely under control, future work should, at least w.r.t the form factors, address the uncertainties on the NLO electroweak corrections. For meaningful experimental studies the existing event generators will have to actively supported.  

{\em The full exploitation of the potential of $t$~quark physics at the ILC requires a very close collaboration between  theoretical and experimental 
groups over the coming years.} The points outlined in this contribution may serve as a basis for the establishment of such a collaboration.